\documentclass{ws-ijmpd}%
\usepackage[super,compress]{cite}%
\begin{document}

\markboth{Remo Garattini}
{Instructions for Typing Manuscripts (Paper's Title)}

%
\catchline{}{}{}{}{}
%

\title{Traversable Wormholes in Distorted Gravity}

\author{Remo Garattini}

\address{Universit\`{a} degli Studi di Bergamo,\\
Dipartimento di Ingegneria e Scienze Applicate,\\
Viale Marconi,5 24044 Dalmine (Bergamo) ITALY\\
I.N.F.N. - sezione di Milano, Milan, Italy.\\
remo.garattini@unibg.it}

\maketitle

\begin{history}
\received{Day Month Year}
\revised{Day Month Year}
\end{history}

\begin{abstract}
We consider the effects of Distorted Gravity on the
traversability of the wormholes. In particular, we consider
configurations which are sustained by their own gravitational
quantum fluctuations. The Ultra-Violet divergences appearing to one loop
are taken under control with the help of a Noncommutative geometry representation 
and Gravity's Rainbow. In this context, it will be shown
that for every framework, the self-sustained equation will produce a Wheeler
wormhole, namely a wormhole of Planckian size. This means that, from the
point of view of traversability, the wormhole will be traversable in
principle, but not in practice. To this purpose, in the context of Gravity's Rainbow
we have considered different proposals of rainbow's functions to see if the smallness
 of the wormhole is dependent on the chosen form of the rainbow's function.
Unfortunately, we discover that this is not the case and we suggest that the
self-sustained equation can be improved to see if the wormhole
radius can be enlarged or not. Some consequences on topology change are discussed.
\end{abstract}


\ccode{04.60.-m, 04.20.Jb}

\maketitle

\section{Introduction}

\label{sec:1}In recent years many attempts to modify gravity have been done
to explain large and short scale phenomena. This modifications seem to be
necessary when one desires to go beyond General Relativity. One proposal
comes from adding higher-order curvature invariants and non-minimally
coupled scalar fields into dynamics resulting from the effective action of
Quantum Gravity \cite{BOS}. Such corrective terms seem to be unavoidable if
we want to obtain the effective action of Quantum Gravity on scales closed
to the Planck length \cite{vilkovisky}. Therefore terms of the form $%
\mathcal{R}^{2}$, $\mathcal{R}^{\mu \nu }\mathcal{R}_{\mu \nu }$, $\mathcal{R%
}^{\mu \nu \alpha \beta }\mathcal{R}_{\mu \nu \alpha \beta }$, $\mathcal{R}%
\,\Box \mathcal{R}$, or $\mathcal{R}\,\Box ^{k}\mathcal{R}$ have to be added
to the effective Lagrangian of gravitational field when quantum corrections
are considered. These higher curvature terms can be included in more general
forms known as $f(\mathcal{R})$ theories and further generalizations%
\footnote{%
For a recent review, see Refs.\cite{SOSO,SCMdL,TSVF}.}. Another proposal comes
from modifying the short scale behavior in an attempt to include
quantum gravitational effects in the description. To this purpose,
Noncommutative geometry, Gravity's Rainbow and Generalized Uncertainty
Principle (GUP) have been widely used to keep under control Ultraviolet
(UV) divergences. For example in a Noncommutative spacetime, one introduces
a commutator $\left[ \mathbf{x}^{\mu },\mathbf{x}^{\nu }\right] =i\,\theta
^{\mu \nu }$, where $\theta ^{\mu \nu }$ is an antisymmetric matrix which
determines the fundamental discretization of spacetime. This feature
eliminates point-like structures in favor of smeared objects in flat
spacetime \cite{Smailagic:2003yb}. Therefore, one may consider the possibility
that noncommutativity could cure the divergences that appear in general
relativity. The effect of the smearing is mathematically implemented with a
substitution of the Dirac-delta function by a Gaussian distribution of
minimal length $\sqrt{\theta }$. In particular, the energy density of a
static and spherically symmetric, smeared and particle-like gravitational
source has been considered in the following form \cite{Nicolini:2005vd}%
\begin{equation}
\rho _{\theta }(r)=\frac{M}{(4\pi \theta )^{3/2}}\;\mathrm{exp}\left( -\frac{%
r^{2}}{4\theta }\right) \,,  \label{NCGenergy}
\end{equation}%
where the mass $M$ is diffused throughout a region of linear dimension $%
\sqrt{\theta }$ due to the intrinsic uncertainty encoded in the coordinate
commutator. On the other hand, when Gravity's Rainbow is considered,
spacetime is endowed with two arbitrary functions $g_{1}\left(
E/E_{P}\right) $ and $g_{2}\left( E/E_{P}\right) $ having the following
properties%
\begin{equation}
\lim_{E/E_{P}\rightarrow 0}g_{1}\left( E/E_{P}\right) =1\qquad \text{and}%
\qquad \lim_{E/E_{P}\rightarrow 0}g_{2}\left( E/E_{P}\right) =1. \label{g1g2}
\end{equation}%
$g_{1}\left( E/E_{P}\right) $ and $g_{2}\left( E/E_{P}\right) $ appear into
the solutions of the modified Einstein's Field Equations\cite{MagSmo}%
\begin{equation}
G_{\mu \nu }\left( E/E_{P}\right) =8\pi G\left( E/E_{P}\right) T_{\mu \nu
}\left( E/E_{P}\right) +g_{\mu \nu }\Lambda \left( E/E_{P}\right) ,
\end{equation}%
where $G\left( E/E_{P}\right) $ is an energy dependent Newton's constant,
defined so that $G\left( 0\right) $ is the low-energy Newton's constant and $%
\Lambda \left( E/E_{P}\right) $ is an energy dependent cosmological
constant. Usually $E$ is the energy associated to the particle deforming
the spacetime geometry. Since the scale of deformation involved is the
Planck scale, it is likely that spacetime itself begins to fluctuate in such a way to
produce a Zero Point Energy (ZPE). In absence of matter fields, the only
particle compatible with the deformed Einstein's gravity is the graviton.
As regards the \textit{Generalized Uncertainty Principle} (GUP)\cite{XL,RenQinChun,GAC1,Mod}
which is a modification of the Heisenberg uncertainty relations, we find that%
\begin{equation}
\Delta x\Delta p\geq \hbar +\frac{\lambda _{p}^{2}}{\hbar }\left( \Delta
p\right) ^{2},
\end{equation}%
where $\hbar $ is the Planck constant and $\lambda _{p}$ is the Planck
length. When the GUP is applied to the Liouville measure, the modified number of quantum
states is%
\begin{equation}
\frac{d^{3}xd^{3}p}{\left( 2\pi \hbar \right) ^{3}\left( 1+\lambda
p^{2}\right) ^{3}}.  \label{eqn:states}
\end{equation}%
For $\lambda =0$, the formula reduces to the ordinary counting of quantum
states. If Eq.$\left( \ref{eqn:states}\right) $ is used for computing black
hole entropy, the usual divergence which appear (brick wall) can be removed%
\cite{XL,RenQinChun}\footnote{%
For applications in Quantum Cosmology, see Ref.\cite{FM}.}. To conclude, one
could also include, as a further example of a theory which takes under control UV divergences,
 a very recent and interesting theory known as Ho\v{r}%
ava-Lifshitz (HL) theory. This is based on a modification of Einstein
gravity motivated by the Lifshitz theory in solid state physics\cite{Horava}%
\cite{Lifshitz}. This modification allows the theory to be power-counting UV
renormalizable and should recover general relativity in the infrared (IR)
limit. Nevertheless, HL theory has the unpleasant feature of being noncovariant. Noncommutative geometry,
Gravity's Rainbow, GUP and HL theory are all examples of \textquotedblleft 
\textit{Distorted Gravity}\textquotedblright . Usually calculations
involving quantum fluctuations of the gravitational field manifest UV
divergences. To keep under control the UV divergences, one invokes a
standard regularization/renormalization process. However in Distorted
Gravity this procedure can be avoided\cite{RemoGRw}, especially when a ZPE calculation is considered.
What makes a ZPE calculation interesting is that it is strictly related to the Casimir
effect. Casimir effect has many applications and it can be considered under different points of view, but it can also be
used as a tool to probe another appealing production of the gravitational field theory: a wormhole.
 A wormhole is often termed Einstein-Rosen bridge because
a \textquotedblleft \textit{bridge}\textquotedblright\ connecting two
\textquotedblleft sheets\textquotedblright\ was the result obtained by A.
Einstein and N. Rosen in attempting to build a geometrical model of a
physical elementary "particle" that was everywhere finite and singularity
free\cite{ER}. It was J.A. Wheeler who introduced the term wormhole\cite{JAW}%
, although his wormholes were at the quantum scale. We have to wait for M.
S. Morris and K. S. Thorne\cite{MT} to see the subject of wormholes
seriously considered by the scientific community. In practice a traversable
wormhole is a solution of the Einstein's Field equations, represented by two
asymptotically flat regions joined by a bridge or, in other word, it is a
short-cut in space and time. To exist, traversable wormholes must violate
the null energy conditions, which means that the matter threading the
wormhole's throat has to be \textquotedblleft \textit{exotic}%
\textquotedblright . Classical matter satisfies the usual energy conditions.
Therefore, it is likely that wormholes must belong to the realm of
semi-classical or perhaps a possible quantum theory of the gravitational
field. On this ground, the Casimir energy on a fixed background. has the
correct properties to substitute the exotic matter: indeed, it is known
that, for different physical systems, Casimir energy is negative. Usually
one considers some matter or gauge fields which contribute to the Casimir
energy necessary to the traversability of the wormholes, nevertheless
nothing forbids to use the Casimir energy of the graviton on a background of
a traversable wormhole\footnote{%
Note that in Ref. \cite{DBGL}, the Casimir energy was used as an indicator
of topology change between wormholes and dark energy stars.}. In this way,
one can think that the quantum fluctuations of the gravitational field of a
traversable wormhole are the same ones which are responsible to sustain
traversability. In this contribution, as an example of Distorted Gravity, we
will fix our attention primarily on Gravity's Rainbow with one extension to 
Noncommutative geometry. The rest of the paper is structured as follows,
in section \ref{p2} we define what is a self-sustained traversable wormhole,
in section \ref{p3} we compute the ZPE graviton energy responsible of the
self sustained traversable wormhole in the context of Distorted Gravity, in
section \ref{p4} we relax the convergence conditions of the integrals
involved in the self-sustained calculation. We summarize and conclude in
section \ref{p5}. Units in which $\hbar =c=k=1$ are used throughout the
paper.

\section{Self-sustained Traversable Wormholes}

\label{p2}In this Section we shall consider the formalism outlined in detail
in Refs. \cite{Remo,Remo1}, where the graviton one loop contribution to a
classical energy in a wormhole background is used. The spacetime metric
representing a spherically symmetric and static wormhole is given by 
\begin{equation}
ds^{2}=-e^{2\Phi (r)}\,dt^{2}+\frac{dr^{2}}{1-b(r)/r}+r^{2}\,(d\theta
^{2}+\sin ^{2}{\theta }\,d\phi ^{2})\,,  \label{metricwormhole}
\end{equation}%
where $\Phi (r)$ and $b(r)$ are arbitrary functions of the radial
coordinate, $r$, denoted as the redshift function, and the shape function,
respectively \cite{MT}. The radial coordinate has a range that increases
from a minimum value at $r_{0}$, corresponding to the wormhole throat, to
infinity. A fundamental property of a wormhole is that a flaring out
condition of the throat, given by $(b-b^{\prime }r)/b^{2}>0$, is imposed 
\cite{MT,Visser}, and at the throat $b(r_{0})=r=r_{0}$, the condition $%
b^{\prime }(r_{0})<1$ is imposed to have wormhole solutions. Another
condition that needs to be satisfied is $1-b(r)/r>0$. For the wormhole to be
traversable, one must demand that there are no horizons present, which are
identified as the surfaces with $e^{2\Phi }\rightarrow 0$, so that $\Phi (r)$
must be finite everywhere. In order to describe a self-sustained traversable
wormhole, we need to define the classical energy. A key point is given by
the background field super-hamiltonian, $\mathcal{H}^{(0)}$. This can be
built with the help of the Arnowitt-Deser-Misner ($\mathcal{ADM}$)
decomposition\cite{ADM} of space time based on the following line element%
\begin{equation}
ds^{2}=g_{\mu \nu }\left( x\right) dx^{\mu }dx^{\nu }=\left(
-N^{2}+N_{i}N^{i}\right) dt^{2}+2N_{j}dtdx^{j}+g_{ij}dx^{i}dx^{j},
\end{equation}%
where $N$ is the \textit{lapse }function and $N_{i}$ the \textit{shift }%
function. In terms of the $\mathcal{ADM}$ variables, the four dimensional
scalar curvature $\mathcal{R}$ can be decomposed in the following way%
\begin{equation}
\mathcal{R}=R+K_{ij}K^{ij}-\left( K\right) ^{2}-2\nabla _{\mu }\left(
Ku^{\mu }+a^{\mu }\right) ,  \label{R}
\end{equation}%
where%
\begin{equation}
K_{ij}=-\frac{1}{2N}\left[ \partial _{t}g_{ij}-N_{i|j}-N_{j|i}\right]
\end{equation}%
is the second fundamental form, $K=$ $g^{ij}K_{ij}$ is its trace, $R$ is the
three dimensional scalar curvature and $\sqrt{g}$ is the three dimensional
determinant of the metric. The last term in $\left( \ref{R}\right) $
represents the boundary terms contribution where the four-velocity $u^{\mu }$
is the timelike unit vector normal to the spacelike hypersurfaces
(t=constant) denoted by $\Sigma _{t}$ and $a^{\mu }=u^{\alpha }\nabla
_{\alpha }u^{\mu }$ is the acceleration of the timelike normal $u^{\mu }$.
Thus%
\begin{equation}
\mathcal{L}\left[ N,N_{i},g_{ij}\right] =\sqrt{-\text{\/\thinspace
\thinspace }^{4}\text{\/{}\negthinspace }g}\ \mathcal{R}=\frac{N}{16\pi G}%
\sqrt{g}\text{ }\left[ K_{ij}K^{ij}-K^{2}+\,R-2\nabla _{\mu }\left( Ku^{\mu
}+a^{\mu }\right) \right]  \label{Lag}
\end{equation}%
represents the gravitational Lagrangian density and $G$ is the Newton's
constant. After a Legendre transformation, we obtain two classical
constraints:%
\begin{equation}
\mathcal{H}^{(0)}=\left( 16\pi G\right) G_{ijkl}\pi ^{ij}\pi ^{kl}-\frac{%
\sqrt{g}}{16\pi G}\!{}\!\,\ R=0  \label{WDWO}
\end{equation}%
and%
\begin{equation}
\mathcal{H}_{j}=\pi _{|j}^{ij}=0.
\end{equation}%
$G_{ijkl}$ is the super-metric and the super-momentum $\pi ^{ij}$ is defined
as%
\begin{equation}
\pi ^{ij}=\frac{\delta \mathcal{L}}{\delta \left( \partial _{t}g_{ij}\right) 
}=\left( g^{ij}K-K^{ij}\text{ }\right) \frac{\sqrt{g}}{16\pi G}.  \label{mom}
\end{equation}%
Note that $\mathcal{H}=0$ represents the classical constraint which
guarantees the invariance under time reparametrization. The other classical
constraint represents the invariance by spatial diffeomorphism and the
vertical stroke \textquotedblleft $|$\textquotedblright\ denotes the
covariant derivative with respect to the $3D$ metric $g_{ij}$. Note that boundary terms
become important when one compares different configurations like Wormholes and Dark Stars\cite{DBGL}
or Wormholes and Gravastars\cite{RemoJHEP}. When we deal
with spherically symmetric line elements, the kinetic term disappears and
the hamiltonian constraint $\left( \ref{WDWO}\right) $ reduces to%
\begin{align}
H_{\Sigma }^{(0)}& =\int_{\Sigma }\,d^{3}x\,\mathcal{H}^{(0)}=-\frac{1}{%
16\pi G}\int_{\Sigma }\,d^{3}x\,\sqrt{g}\,R\,  \nonumber \\
& =-\frac{1}{2G}\int_{r_{0}}^{\infty }\,\frac{dr\,r^{2}}{\sqrt{1-b(r)/r}}\,%
\frac{b^{\prime }(r)}{r^{2}}\,,  \label{classical}
\end{align}%
where we have integrated on a constant time hypersurface $\mathcal{H}^{(0)}$
and where we have used the explicit expression of the scalar curvature in
three dimensions in terms of the shape function. Therefore, from the Hamiltonian point of view,
it is not necessary to assume that $\Phi \left( r\right) =const.$. A traversable wormhole is
said to be \textquotedblleft \textit{self sustained}\textquotedblright\ if%
\begin{equation}
H_{\Sigma }^{(0)}=-E^{TT},  \label{SS}
\end{equation}%
where $E^{TT}$ is the total regularized graviton one loop energy. Basically
this is given by 
\begin{equation}
E^{TT}=-\frac{1}{2}\sum_{\tau }\left[ \sqrt{E_{1}^{2}\left( \tau \right) }+%
\sqrt{E_{2}^{2}\left( \tau \right) }\right] \,,  \label{OneL}
\end{equation}%
where $\tau $ denotes a complete set of indices and $E_{i}^{2}\left( \tau
\right) >0$, $i=1,2$ are the eigenvalues of the modified Lichnerowicz
operator%
\begin{equation}
\left( \hat{\bigtriangleup}_{L\!}^{m}\!{}\;h^{\bot }\right) _{ij}=\left(
\bigtriangleup _{L\!}\!{}\;h^{\bot }\right)
_{ij}-4R{}_{i}^{k}\!{}\;h_{kj}^{\bot }+\text{ }^{3}R{}\!{}\;h_{ij}^{\bot }\,,
\end{equation}%
acting on traceless-transverse tensors of the perturbation and where $%
\bigtriangleup _{L}$is the Lichnerowicz operator defined by%
\begin{equation}
\left( \bigtriangleup _{L}\;h\right) _{ij}=\bigtriangleup
h_{ij}-2R_{ikjl}\,h^{kl}+R_{ik}\,h_{j}^{k}+R_{jk}\,h_{i}^{k},
\end{equation}%
with $\bigtriangleup =-\nabla ^{a}\nabla _{a}$. For the background $\left( %
\ref{metricwormhole}\right) $, one can define two r-dependent radial wave
numbers%
\begin{equation}
k_{i}^{2}\left( r,l,\omega _{i,nl}\right) =\omega _{i,nl}^{2}-\frac{l\left(
l+1\right) }{r^{2}}-m_{i}^{2}\left( r\right) \quad i=1,2\quad ,  \label{kTT}
\end{equation}%
where%
\begin{equation}
\left\{ 
\begin{array}{c}
m_{1}^{2}\left( r\right) =\frac{6}{r^{2}}\left( 1-\frac{b\left( r\right) }{r}%
\right) +\frac{3}{2r^{2}}b^{\prime }\left( r\right) -\frac{3}{2r^{3}}b\left(
r\right) \\ 
\\ 
m_{2}^{2}\left( r\right) =\frac{6}{r^{2}}\left( 1-\frac{b\left( r\right) }{r}%
\right) +\frac{1}{2r^{2}}b^{\prime }\left( r\right) +\frac{3}{2r^{3}}b\left(
r\right)%
\end{array}%
\right.  \label{masses}
\end{equation}%
are two r-dependent effective masses $m_{1}^{2}\left( r\right) $ and $%
m_{2}^{2}\left( r\right) $. When we perform the sum over all modes, $E^{TT}$
is usually divergent. In Refs. \cite{Remo,Remo1} a standard
regularization/renormalization scheme has been adopted to handle the
divergences. In this contribution, we will consider the effect of Gravity's
Rainbow and Noncommutative geometry on the graviton to one loop. One
advantage in using such a scheme is to avoid the renormalization process and
to use only one scale: the Planck scale.

\section{Some examples of Distorted Gravity}

\label{p3}One of the purposes of Eq.$\left( \ref{SS}\right) $ is the
possible discovery of a traversable wormhole with the determination of the
form of the shape function. However, one could also reverse the strategy: we
fix the wormhole shape to see if the self-sustained equation is satisfied.
One good candidate is%
\begin{equation}
b\left( r\right) =r_{0}^{2}/r,  \label{b(r)}
\end{equation}%
which is the prototype of the traversable wormholes\cite{Ellis,MT}. Plugging
the shape function $\left( \ref{b(r)}\right) $ into Eq.$\left( \ref{SS}%
\right) $, we find that the left hand side becomes%
\begin{equation}
H_{\Sigma }^{(0)}=\frac{1}{2G}\int_{r_{0}}^{\infty }\,\frac{dr\,r^{2}}{\sqrt{%
1-r_{0}^{2}/r^{2}}}\,\frac{r_{0}^{2}}{r^{4}}\,,
\end{equation}%
while the right hand side is divergent. To handle with divergences, we have
several possibilities. For example, in Ref.\cite{Remo} a
regularization/renormalization scheme has been adopted and the wormhole
throat has been fixed at the value $r_{0}\simeq 1.16/E_{P}$. If we adopt a
Noncommutative scheme\cite{RGPN}, the distorted Liouville measure%
\begin{equation}
dn_{i}=\frac{d^{3}\vec{x}d^{3}\vec{k}}{\left( 2\pi \right) ^{3}}\exp \left( -%
\frac{\theta }{4}\left( \omega _{i,nl}^{2}-m_{i}^{2}\left( r\right) \right)
\right) ,\quad i=1,2,  \label{dn}
\end{equation}%
allows the computation of the graviton to one loop. This is possible because
the distortion induced by the Noncommutative space time allows the right
hand side of Eq.$\left( \ref{SS}\right) $ to be finite. Indeed, plugging $%
dn_{i}$ into Eq.$\left( \ref{SS}\right) $, one finds that the self-sustained
equation for the energy density becomes%
\begin{equation}
\frac{3\pi ^{2}}{Gr_{0}^{2}}=\int_{0}^{+\infty }\sqrt{\left( \omega ^{2}+%
\frac{3}{r_{0}^{2}}\right) ^{3}}e^{-\frac{\theta }{4}\left( \omega ^{2}+%
\frac{3}{r_{0}^{2}}\right) }d\omega +\int_{1/r_{0}}^{+\infty }\sqrt{\left(
\omega ^{2}-\frac{1}{r_{0}^{2}}\right) ^{3}}e^{-\frac{\theta }{4}\left(
\omega ^{2}-\frac{1}{r_{0}^{2}}\right) }d\omega ,  \label{Schw1loop}
\end{equation}%
where we have used the shape function $\left( \ref{b(r)}\right) $ to
evaluate the effective masses $\left( \ref{masses}\right) $. By defining the
dimensionless variable%
\begin{equation}
x=\frac{\theta }{4r_{0}^{2}},
\end{equation}%
Eq.$\left( \ref{Schw1loop}\right) $ leads to $\left( G=l_{P}^{2}\right) $%
\[
\frac{3\pi ^{2}\theta }{l_{P}^{2}}=F\left( x\right) , 
\]%
where%
\begin{gather}
F\left( x\right) =\left( \left( 1-x\right) K_{1}\left( \frac{x}{2}\right)
+xK_{0}\left( \frac{x}{2}\right) \right) \exp \left( \frac{x}{2}\right) 
\nonumber \\
+3\left( \left( 1+3x\right) K_{1}\left( \frac{3x}{2}\right) +3xK_{0}\left( 
\frac{3x}{2}\right) \right) \exp \left( -\frac{3x}{2}\right) .
\end{gather}%
$F\left( x\right) $ has a maximum for $\bar{x}=0.24$, where%
\begin{equation}
\frac{3\pi ^{2}\theta }{l_{P}^{2}}=F\left( \bar{x}\right) =2.20.
\end{equation}%
This fixes $\theta $ to be%
\begin{equation}
\theta =\frac{2.20l_{P}^{2}}{3\pi ^{2}}=7.\,\allowbreak 43\times
10^{-2}l_{P}^{2}.  \label{theta}
\end{equation}%
and%
\begin{equation}
r_{0}=0.28l_{P}.
\end{equation}%
When we consider Gravity's
Rainbow, spacetime is endowed with two arbitrary functions $g_{1}\left(
E/E_{P}\right) $ and $g_{2}\left( E/E_{P}\right) $ having the properties shown in $\left( \ref{g1g2}\right) $.
 As shown in Ref.\cite{RGFSNL}, the self-sustained equation $\left( \ref{SS}\right) $
becomes 
\begin{equation}
\frac{b^{\prime }(r)}{2Gg_{2}\left( E/E_{P}\right) r^{2}}=\frac{2}{3\pi ^{2}}%
\left( I_{1}+I_{2}\right) \,,  \label{ETT}
\end{equation}%
where the r.h.s. of Eq.$\left( \ref{ETT}\right) $ is represented by%
\begin{equation}
I_{1}=\int_{E^{\ast }}^{\infty }E\frac{g_{1}\left( E/E_{P}\right) }{%
g_{2}^{2}\left( E/E_{P}\right) }\frac{d}{dE}\left( \frac{E^{2}}{%
g_{2}^{2}\left( E/E_{P}\right) }-m_{1}^{2}\left( r\right) \right) ^{\frac{3}{%
2}}dE\,  \label{I1}
\end{equation}%
and%
\begin{equation}
I_{2}=\int_{E^{\ast }}^{\infty }E\frac{g_{1}\left( E/E_{P}\right) }{%
g_{2}^{2}\left( E/E_{P}\right) }\frac{d}{dE}\left( \frac{E^{2}}{%
g_{2}^{2}\left( E/E_{P}\right) }-m_{2}^{2}\left( r\right) \right) ^{\frac{3}{%
2}}dE\,,  \label{I2}
\end{equation}%
respectively. $E^{\ast }$ is the value which annihilates the argument of the
root. Of course, $I_{1}$ and $I_{2}$ are finite for appropriate choices of
the Rainbow's functions $g_{1}\left( E/E_{P}\right) $ and $g_{2}\left(
E/E_{P}\right) $. If we assume that%
\begin{equation}
g_{1}\left( E/E_{P}\right) =\exp \left( -\alpha E^{2}/E_{P}^{2}\right)
\qquad g_{2}\left( E/E_{P}\right) =1\,,  \label{GRF}
\end{equation}%
with $\alpha \in 
\mathbb{R}
$, the classical term is not distorted. We find%
\begin{equation}
I_{1}=3\int_{\sqrt{3/r_{0}^{2}}}^{\infty }\exp \left( -\alpha
E^{2}/E_{P}^{2}\right) E^{2}\sqrt{E^{2}-\frac{3}{r_{0}^{2}}}dE
\end{equation}%
and%
\begin{equation}
I_{2}=3\int_{0}^{\infty }\exp \left( -\alpha E^{2}/E_{P}^{2}\right) E^{2}%
\sqrt{E^{2}+\frac{1}{r_{0}^{2}}}dE\,,
\end{equation}%
where we have fixed the shape function as in Eq.$\left( \ref{b(r)}\right) $.
Now, in order to have only one solution with variables $\alpha $ and $r_{0}$%
, we demand that%
\begin{equation}
\frac{d}{dr_{0}}\left[ -\frac{1}{2G}\,\frac{1}{r_{0}^{2}}\right] \,=\frac{d}{%
dr_{0}}\left[ \frac{2}{3\pi ^{2}}\left( I_{1}+I_{2}\right) \right] \,,
\end{equation}%
which takes the following form after the integration $\left(
G^{-1}=E_{P}^{2}\right) $%
\begin{equation}
1=\frac{1}{2\pi ^{2}x^{2}}f\left( \alpha ,x\right) \,,  \label{SSEq1}
\end{equation}%
where $x=r_{0}E_{P}$ and where 
\begin{gather}
f\left( \alpha ,x\right) =\exp \left( \frac{\alpha }{2x^{2}}\right)
K_{0}\left( \frac{\alpha }{2x^{2}}\right) -\exp \left( \frac{\alpha }{2x^{2}}%
\right) K_{1}\left( \frac{\alpha }{2x^{2}}\right)  \nonumber \\
+9\exp \left( -\frac{3\alpha }{2x^{2}}\right) K_{0}\left( \frac{3\alpha }{%
2x^{2}}\right) +\exp \left( -\frac{3\alpha }{2x^{2}}\right) K_{1}\left( 
\frac{3\alpha }{2x^{2}}\right) .
\end{gather}%
$K_{0}(x)$ and $K_{1}(x)$ are the modified Bessel function of order 0 and 1,
respectively. One finds a root at%
\begin{equation}
\bar{x}=r_{0}E_{P}=2.973786871\sqrt{\alpha }  \label{xmin}
\end{equation}%
with $\alpha \simeq 0.242$. This means that $r_{0}E_{P}=\allowbreak
1.\,\allowbreak 46$. It is immediate to understand that the example coming
from Noncommutative geometry appears to be \textquotedblleft \textit{rigid}%
\textquotedblright . With this term we mean that the distorted Liouville
measure cannot assume different forms. On the other hand, Gravity's Rainbow
depends on the choice of $g_{1}\left( E/E_{P}\right) $ and $g_{2}\left(
E/E_{P}\right) $. In this short presentation we have explored only the
Gaussian proposal $\left( \ref{GRF}\right) $ which strongly cuts off the
Planckian and trans-Planckian physics. To this purpose, we would like to
considerably relax the convergent form of the rainbow's functions to see if there are some
effects on the traversability of the wormhole.

\section{Relaxing Gravity's Rainbow}

\label{p4}In this part of the contribution, we would like to explore a
relaxed variant of the choice $\left( \ref{GRF}\right) $. The situation we
are going to consider is the following. It is easy to see that if we assume%
\begin{equation}
g_{1}\left( E/E_{P}\right) =1\qquad g_{2}\left( E/E_{P}\right) =\left\{ 
\begin{array}{c}
1\qquad \mathrm{when}\qquad E<E_{P} \\ 
\\ 
E/E_{P}\qquad \mathrm{when}\qquad E>E_{P}\qquad%
\end{array}%
\right. \,,  \label{rel}
\end{equation}%
$\left( \ref{I1}\right) $ and $\left( \ref{I2}\right) $ reduce to%
\begin{equation}
I_{1}=3\int_{\sqrt{m_{1}^{2}\left( r\right) }}^{E_{P}}E^{2}\sqrt{%
E^{2}-m_{1}^{2}\left( r\right) }dE\,,
\end{equation}%
and%
\begin{equation}
I_{2}=3\int_{\sqrt{m_{2}^{2}\left( r\right) }}^{E_{P}}E^{2}\sqrt{%
E^{2}-m_{2}^{2}\left( r\right) }dE\,.
\end{equation}%
As a specific choice for the shape function we will take under consideration
old examples suggested by Morris and Thorne\cite{MT}:%
\begin{equation}
b\left( r\right) =\frac{4}{3}r_{0}-\frac{1}{3}r_{0}\left( \frac{r_{0}}{r}%
\right) ^{2}  \label{b(r)a}
\end{equation}%
and%
\begin{equation}
b\left( r\right) =\sqrt{r_{0}r}.  \label{b(r)b}
\end{equation}%
For both of them, we will also assume that $\Phi \left( r\right) =const.$ %
Both the examples $\left( \ref{b(r)a}\right) $ and $\left( \ref{b(r)b}%
\right) $ satisfy%
\begin{equation}
b\left( r_{0}\right) =r_{0},\qquad \frac{b\left( r\right) }{r}\rightarrow
0\qquad \mathrm{when}\qquad r\rightarrow \infty .
\end{equation}%
As regards the choice $\left( \ref{b(r)a}\right) $, one finds%
\begin{equation}
b^{\prime }\left( r\right) =\frac{2}{3}\left( \frac{r_{0}}{r}\right) ^{3}<1
\label{b'(r)a}
\end{equation}%
and the flare out condition is satisfied on the throat. To see if the
wormhole can be self-sustained, we plug the shape function $\left( \ref%
{b(r)a}\right) $ into $\left( \ref{I1}\right) $ and $\left( \ref{I2}\right) $%
. Then Eq. $\left( \ref{ETT}\right) $ simply becomes%
\begin{equation}
\frac{r_{0}^{3}}{3Gr^{5}}=\frac{2}{\pi ^{2}}\left( \int_{\sqrt{%
m_{1}^{2}\left( r_{0}\right) }}^{E_{P}}E^{2}\sqrt{E^{2}-m_{1}^{2}\left(
r_{0}\right) }dE+\int_{\sqrt{m_{2}^{2}\left( r_{0}\right) }}^{E_{P}}E^{2}%
\sqrt{E^{2}-m_{2}^{2}\left( r_{0}\right) }dE\right) \,,  \label{ETT1}
\end{equation}%
where we have explicitly used the form of the shape function $\left( \ref%
{b(r)a}\right) $ on the classical term. On the throat, the effective masses
become%
\begin{equation}
m_{1}^{2}\left( r_{0}\right) =\frac{1}{2r_{0}^{2}}\qquad m_{2}^{2}\left(
r_{0}\right) =-\frac{11}{6r_{0}^{2}}
\end{equation}%
and Eq.$\left( \ref{ETT1}\right) $ simplifies into%
\begin{equation}
\frac{E_{P}^{2}}{3r_{0}^{2}}=\frac{2}{\pi ^{2}}\left( \int_{\sqrt{%
1/2r_{0}^{2}}}^{E_{P}}E^{2}\sqrt{E^{2}-\frac{1}{2r_{0}^{2}}}dE+\int_{\sqrt{%
11/6r_{0}^{2}}}^{E_{P}}E^{2}\sqrt{E^{2}+\frac{11}{6r_{0}^{2}}}dE\right) \,.
\label{ETT2}
\end{equation}%
The explicit calculation of $\left( \ref{ETT2}\right) $ leads to

\begin{gather}
1=\frac{1}{\pi^{2}}\left[ \frac{3x^{2}}{2}\left( 1-\frac{1}{2x^{2}}\right)
^{3/2}+{\frac{3}{8}\sqrt{1-\frac{1}{2x^{2}}}}-{\frac{3}{16\,x^{2}}\ln\left( 
\sqrt{2}x+\sqrt{4x^{2}-1}\right) }\right.  \nonumber \\
\left. +\frac{3x^{2}}{2}\left( 1+{\frac{11}{6\,x^{2}}}\right) ^{3/2}-\frac{11%
}{8}\sqrt{1+{\frac{11}{6x^{2}}}}-\frac{121}{48x^{2}}{\ln\left( \sqrt{\frac{%
6x^{2}}{11}}+\sqrt{\frac{6x^{2}}{11}+{1}}\right) }\right] ,
\end{gather}
where $x={r}_{0}E_{P}$. The solution can be easily computed numerically and
we find $x={r}_{0}E_{P}=1.70$. As regards the choice $\left( \ref{b(r)b}%
\right) $, one finds%
\begin{equation}
b^{\prime}\left( r\right) =\frac{1}{2}\sqrt{\frac{r_{0}}{r}}<1
\end{equation}
and, once again, the flare out condition is satisfied on the throat. In this
case, the self-sustained equation reduces to%
\begin{equation}
\frac{r_{0}}{4Gr^{3}}=\frac{2}{\pi^{2}}\left( \int_{\sqrt{m_{1}^{2}\left(
r_{0}\right) }}^{E_{P}}E^{2}\sqrt{E^{2}-m_{1}^{2}\left( r_{0}\right) }%
dE+\int_{\sqrt{m_{2}^{2}\left( r_{0}\right) }}^{E_{P}}E^{2}\sqrt {%
E^{2}-m_{2}^{2}\left( r_{0}\right) }dE\right) \,.  \label{ETT3}
\end{equation}
The effective masses become on the throat%
\begin{equation}
m_{1}^{2}\left( r_{0}\right) =\frac{3}{4r_{0}^{2}}\qquad m_{2}^{2}\left(
r_{0}\right) =-\frac{7}{4r_{0}^{2}}
\end{equation}
and Eq.$\left( \ref{ETT3}\right) $ simplifies into%
\begin{equation}
\frac{E_{P}^{2}}{3r_{0}^{2}}=\frac{2}{\pi^{2}}\left( \int_{\sqrt{3/4r_{0}^{2}%
}}^{E_{P}}E^{2}\sqrt{E^{2}-\frac{3}{4r_{0}^{2}}}dE+\int_{\sqrt {7/4r_{0}^{2}}%
}^{E_{P}}E^{2}\sqrt{E^{2}+\frac{7}{4r_{0}^{2}}}dE\right) \,.  \label{ETT4}
\end{equation}
The explicit calculation of $\left( \ref{ETT4}\right) $ leads to

\begin{gather}
1=\frac{1}{\pi^{2}}\left[ 2x^{2}\left( 1-\frac{3}{4x^{2}}\right) ^{3/2}+{%
\frac{3}{4}\sqrt{1-\frac{3}{4x^{2}}}}-{\frac{9}{16x^{2}}\ln\left( \sqrt{%
\frac{4x^{2}}{3}}+\sqrt{\frac{4x^{2}}{3}-1}\right) }\right.  \nonumber \\
\left. 2x^{2}\left( 1+\frac{7}{4x^{2}}\right) ^{3/2}-{\frac{7}{4}\sqrt{1+%
\frac{7}{4x^{2}}}}-{\frac{49}{16\,x^{2}}\ln\left( \sqrt{\frac{4x^{2}}{7}}+%
\sqrt{\frac{4x^{2}}{7}+1}\right) }\right] ,
\end{gather}
where $x={r}_{0}E_{P}$. Even in this case, the solution can be easily
computed numerically and we find $x={r}_{0}E_{P}=1.51$. Note that the form
in $\left( \ref{b(r)b}\right) $ is a special case of a shape function of the
form%
\begin{equation}
b\left( r\right) =r_{0}\left( \frac{r_{0}}{r}\right) ^{\omega},
\end{equation}
which is obtained imposing an equation of state $p_{r}=\omega\rho$.

\section{Conclusions}

\label{p5}In this contribution we have discussed how some forms of Distorted
Gravity can be used to power a traversable wormhole. The distorted Liouville
measure $\left( \ref{dn}\right) $ has been used as an example of
Noncommutative geometry. We have obtained a solution compatible with the
procedure located at $r_{0}E_{P}=0.28$\cite{RemoTM}. Note that in Ref.\cite%
{RGFSNLPLB}, a shape form induced by a density profile of the form $\left( %
\ref{NCGenergy}\right) $ but with an ordinary regularization/renormalization
procedure. It is beyond this contribution to investigate the shape function
obtained by $\left( \ref{NCGenergy}\right) $ combined with the distorted
Liouville measure $\left( \ref{dn}\right) $ gives a traversable wormhole. As
regards Gravity's Rainbow; the good news is that every shape function
analyzed is traversable. The bad news is that the traversability is in
principle but not in practice, even if the radii are greater that the radius
discovered in Ref.\cite{Remo} and also larger than the one obtained with the
measure $\left( \ref{dn}\right) $. Indeed, all the wormholes discovered
survive in the Planckian or Trans-Planckian regime. This is because we are
probing a region where the gravitational field develops quantum fluctuations
so violent that it is also able to give a topology change\cite{EPJC}. Since
the wormhole's radius is of the Planckian size, we have explored a relaxed
version of the rainbow's functions to see if the smallness of the wormhole
radius is a consequence of the strong convergence induced by the choice $%
\left( \ref{GRF}\right) $. Once again, the resulting wormhole is traversable
in principle but not in practice. This means that we can interpret the
self-sustained equation as an ignition equation. To obtain a larger radius,
one possibility is to use the self-sustained equation in the following manner%
\begin{equation}
\frac{1}{2G}\,\frac{\left( b^{\prime }(r)\right) ^{\left( n\right) }}{%
r^{2}g_{2}\left( E\right) }=\frac{2}{3\pi ^{2}}\left[ I_{1}\left( b^{\left(
n-1\right) }(r)\right) +I_{2}\left( b^{\left( n-1\right) }(r)\right) \right]
,  \label{LoG}
\end{equation}%
where $n$ is the order of the approximation. In this way, if we discover
that fixing the radius to some value of a fixed background of the r.h.s.,
and we discover on the l.h.s. a different radius, we could conclude that if
the radius is larger that the original, the wormhole is growing, otherwise
is collapsing. Note that in Ref.\cite{EPJC}, Eq. $\left( \ref{LoG}\right) $
has been used to show that a traversable wormhole can be generated with a
topology change starting from a Minkowski spacetime.

\end{document}